\documentclass{aa}
\usepackage{graphicx}
\usepackage{txfonts}
\usepackage{lscape}
\usepackage{natbib}

\begin{document}
\title{The changing look of PKS 2149-306}

\author{V.~Bianchin\inst{1} \and L.~Foschini\inst{1} \and G.~Ghisellini\inst{2} \and G.~Tagliaferri\inst{2} \and F.~Tavecchio\inst{2} \and A.~Treves\inst{3} \and G.~Di~Cocco\inst{1} \and M.~Gliozzi\inst{4} \and E.~Pian\inst{5} \and R.~M.~Sambruna\inst{6} \and A.~Wolter\inst{2} }


\institute{INAF/IASF-Bologna, Via Gobetti 101 - 40129 Bologna, Italy\\\email{bianchin@iasfbo.inaf.it}
\and INAF-Osservatorio Astronomico di Brera, via E. Bianchi 46, 23807 Merate, Italy 
\and Dipartimento di Scienze, Universit\`{a} dell'Insubria, Como, Italy
\and George Mason University, 4400 University Drive, Fairfax, Va 22030, USA 
\and INAF-Osservatorio Astronomico di Trieste, via G.~B.~Tiepolo 11, 34131 Trieste, Italy 
\and NASA Goddard Space Flight Center, Code 661, Greenbelt, MD 20771, USA}

\date{Received --; accepted --}

 \abstract
  {}
{ We study the blazar nature of the high-redshift Flat-Spectrum Radio 
Quasar PKS~$2149-306$ ($z = 2.345$) by investigating its long-term behavior.}
  {We analyzed all publicly available optical-to-X-ray
observations performed by \emph{XMM-Newton}, \emph{Swift}, and 
\emph{INTEGRAL}.}
  {PKS~$2149-306$ is one of four blazars at $z>2$ that have been observed in 
    the hard-X-ray regime with both the BAT and ISGRI instruments. 
    Observations acquired almost $1$~year apart in the $60-300$~keV energy 
    band in the object rest frame, exhibit no noticeable change in spectral 
    slope associated with a flux variation of more than a factor of two. 
    \emph{Swift} data appear to show a roll-off below $\sim 1$~keV, which 
    becomes increasingly evident during a $\sim 3$-day time-frame, that can 
    be explained as the natural spectral break caused by the Inverse Compton 
    onset. 
    The broad-band spectra allow us to identify two different states. 
    The SED modeling suggests that they can be interpreted by only a change 
    in the bulk Lorentz factor of the jet.}
  {}

\authorrunning{V. Bianchin et al.}

\maketitle

\section{Introduction}
PKS~$2149-306$ is a Flat Spectrum Radio Quasar (FSRQ) located at $z=2.345$ 
\citep{bw86}. 
In the radio maps, it exhibits a compact but non-point-like structure 
(e.g.,~Ojha et al.~2005) of $\approx 1$~Jy flux density over a wide 
radio frequency range ($0.8-8.4$~GHz)\footnote{Data from NASA/IPAC Extragalactic Database (NED)  \texttt{http://nedwww.ipac.caltech.edu/index.html}}. 
The few observations of the optical counterpart show no significant changes in 
all filters with $m_V \sim 18$ (e.g.,~Francis et al.~2000). 
The object hosts a bright X-ray source extensively observed in the past by 
\emph{ROSAT} and \emph{ASCA} \citep{SIEBERT96,CAPPI97}, 
\emph{BeppoSAX} \citep{ELVIS00}, \emph{Chandra} \citep{FANG01}, 
\emph{XMM-Newton} \citep{FERRERO03}, and more recently by \emph{Swift} 
\citep{SAMBRUNA07}. 
In the X-ray domain, the source exhibits a hard spectrum with variable photon 
index and flux. 
From early \emph{ASCA} observations, the presence of an emission line at 
$\approx 17$~keV in the blazar frame was interpreted as a highly-blueshifted 
Fe~K$\alpha$ line produced by an outflow at speed $v \sim 0.7 c$ 
\citep{YAQOOB99}; however, \emph{Chandra} high-resolution spectroscopy did not 
confirm this finding \citep{FANG01}. 
Page et al.~(2004) found no evidence for a disk/torus Fe~K$\alpha$ emission 
line. 

As for other high-$z$ FSRQs, a controversial low-energy photon deficit has 
been claimed by some authors but not confirmed by others 
\citep{SIEBERT96,CAPPI97,YAQOOB99,SAMBRUNA07,ELVIS00,FERRERO03,PAGE05}.
This putative deficit was usually ascribed to an absorbing cloud in the 
quasar reference frame and modeled by additional intrinsic absorption in 
addition to the Galactic one. 
However, considering the blazar nature of PKS~$2149-306$, it is difficult to
explain how an absorbing cloud can survive in front of a jet. 
On the other hand, interpreting this flux deficit as a natural break in 
the Spectral Energy Distribution (SED) due to the low-energy tail of the 
electron population leads to a coherent explanation of the observed spectral 
changes in PKS~$2149-306$ (cf.~the case of RBS~$315$, Tavecchio et al.~2007).

PKS~$2149-306$ is one of the few FSRQs observed in the hard-X-ray regime after
the \emph{BeppoSAX} era \citep{ELVIS00}.
A recent \emph{Swift} observation \citep{SAMBRUNA07} confirmed that the 
source displays a Compton-dominated spectrum, typical of FSRQs, with a Compton 
peak probably located at approximately hundreds of keV (observer frame). 

In this work, we present a serendipitous detection of PKS~$2149-306$ in 
\emph{INTEGRAL} archival data. 
To date, the source catalog of the \emph{INTEGRAL}/ISGRI detector 
\citep{LEBRUN03} includes $18$ blazars, among which $6$ are located at $z>2$ 
(e.g., Bird et al.~2007). 
The AGN catalog of the \emph{Swift}/BAT instrument \citep{BAT} contains 
comparable numbers, with $6$ high-redshift objects \citep{BATSURVEY}. 
Including PKS~$2149-306$, there are only four objects detected by both ISGRI 
and BAT. 
We performed a spectral analysis of \emph{INTEGRAL} data reanalyzed all 
publicly available data for the source from the \emph{XMM-Newton} and 
\emph{Swift} archives. 
Here, we discuss the broad-band spectral variation of the source over several 
years. 

\section{Data analysis}
\begin{table*}[!t]
\begin{minipage}[t]{\textwidth}
\caption{\label{log} Observation Log.}
\begin{center}
\renewcommand{\footnoterule}{}
\begin{tabular}{llcc}
\hline
\hline
 Instrument     & ObsID        & Start Date   & Exposure \\
                &              & [YYYY-MM-DD] & [ks]     \\
\hline
\emph{INTEGRAL}/ISGRI     & Rev.~$254$, $258-260$ & $2004-11-11$ & $366$    \\
\emph{XMM}/EPIC           & $0103060401$  & $2001-05-01$ & $20$       \\
\emph{Swift}/XRT (OBS.~1) & $00035242001$ & $2005-12-10$ & $3.1$      \\
\emph{Swift}/XRT (OBS.~2) & $00035242002$ & $2005-12-13$ & $2.3$      \\ 
\emph{Swift}/BAT          & 9-month Survey&mid Dec $2005$& $2800$     \\
\hline
\end{tabular}
\end{center}
\footnotesize{For each instrument, the table indicates the observation ID or the revolution number, the starting date of the observation, and the effective exposure for PKS~$2149-306$. BAT data refer to the $9-$month survey, starting in mid-December $2005$. BAT exposure time is computed by integrating over all pointings with the source PKS~$2149-306$ within the FOV of the BAT instrument.} 
\end{minipage}
\end{table*}

Table~\ref{log} lists all observations of PKS~$2149-306$ analyzed in this 
paper, including starting dates and exposure times. 
For all observatories, data were processed using the latest versions of the 
specific software and calibration files. 
The output spectra of \emph{XMM-Newton}/EPIC and \emph{Swift}/XRT were 
rebinned to contain at least $30$ counts per bin. 
Spectral analysis was carried out by means of \texttt{xspec v.12.4.0}. 
In all X-ray model fits, we included Galactic extinction and fixed the 
column density to be $N_{\rm H}=1.63\times 10^{20}$~cm$^{-2}$ 
\citep{LAB}. 
Our fit results are given in Table~\ref{fits}. 

\subsection{INTEGRAL}
We achieved a serendipitous detection of PKS~$2149-306$ using the 
\emph{INTEGRAL} public archive data; the source was in the IBIS 
\citep{UBERTINI03} field of view during November $2004$ (Revolutions 
$254$, $258-260$), while pointing to NGC~$7172$. 
Data were analyzed with the Off-line Scientific analysis Software
\texttt{OSA 7.0} \citep{COURVOISIER}. 
The imaging analysis of IBIS/ISGRI data \citep{GOLDWURM03} provided 
a source detection of signal-to-noise ratio S/N$=4$ in the $20-100$~keV 
energy band with exposure of $366$~ks.
The faint detections in single pointings did not allow us to inspect light 
curve variations. 
The total spectrum was extracted using a rebinned response matrix with $5$ 
energy bins in the energy range $20-140$~keV. 
The ISGRI spectrum can be modeled by a single power-law with photon index
$\Gamma = 1.5^{+0.9}_{-0.8}$ and flux 
$(2.1 \pm 0.2)\times 10^{-11}$~erg~cm$^{-2}$~s$^{-1}$ in the $20-100$~keV 
energy band (Table~\ref{fits}). 

The source is below the capabilities of both JEM-X and PICsIT instruments,
with upper limits of $10^{-11}$~erg~cm$^{-2}$~s$^{-1}$ 
in $8-14$~keV and $3.5 \times 10^{-10}$~erg~cm$^{-2}$~s$^{-1}$ 
in $252-336$~keV respectively. 

SPI data were not considered since the source NGC~$7172$, 
the primary target of the observations, falls on the border of the instrument 
Point-Spread Function (PSF) ($2.5$~degrees) centered on PKS~$2149-306$. 

OMC data were not included in the analysis because a contaminating 
source is located within the instrumental $25$~arcsec PSF FWHM of 
PKS~$2149-306$ (see the \emph{Swift}/UVOT data analysis below).

\subsection{XMM-Newton}
\emph{XMM-Newton} observed PKS~$2149-306$ on May $1$, $2001$, starting at
$10$:$53$~UT (ObsID $0103060401$). 
We reanalyzed these data, already presented by Ferrero \& Brinkmann~(2003). 
Data from EPIC-PN \citep{EPICPN}, EPIC-MOS \citep{MOS} and OM \citep{OM} were 
processed, screened, and analyzed using the same procedure described in 
Foschini et al.~(2006), but with the \texttt{SAS v 7.1.0} software and 
calibration file release of July $16$, $2007$. 
No periods of high-background were detected, and the net exposures with the 
individual detectors were $20.3$~ks with PN and $24$~ks with MOS1 and MOS2. 

The EPIC-PN light curve (not shown here) displays no significant flux changes
($\chi^2$ probability of constancy $0.43$) with an upper limit to the 
fractional variability $<3$\% ($3\sigma$).

Our spectral analysis confirms the results of Ferrero \& Brinkmann~(2003) 
that the best-fit model was a single power-law function with parameters 
reported in Table~\ref{fits}. 
The normalization constants between the MOS instruments and PN are
consistent with unity ($C_{MOS1-PN} = 1.00\pm 0.02$, $C_{MOS2-PN} =
1.03\pm 0.02$). 
In panel \textit{a} of Fig.~\ref{res}, the residuals of the power-law model 
are plotted in units of $\sigma$ and with error bars of size $1~\sigma$; for 
a clearer visualization, only PN data are shown. 
We applied more complex models: a power-law function with Galactic extinction 
and an absorption excess in the source frame (\texttt{wabs*zwabs*zpo}) and 
an absorbed broken power-law function (\texttt{wabs*bknpo}). 
Both provided comparable values of $\chi^2_r$, but the parameters were not 
well-constrained or they converged to unphysical quantities.

We also checked for the presence of neutral iron line emission in the quasar 
frame but we found no indication of such a feature (confirming 
Page et al.~2004). 

The OM observed the blazar for $\sim 1000$~s with $UVW2$ filter (centered at 
$212$~nm): no detection was found, with a magnitude lower limit of 
$19$ ($3\sigma$), corresponding to a flux 
$\le 3.4 \times 10^{-13}$~erg~cm$^{-2}$~s$^{-1}$. 

\subsection{Swift}
\emph{Swift} observed PKS~$2149-306$ twice in $2005$ \citep{SAMBRUNA07}: 
ObsID $00035242001$ started on December $10$, $2005$ $00$:$47$~UT, and 
ObsID $00035242002$ started  on December $13$, $2005$ $20$:$13$~UT with 
exposure times of $3.1$~ks and $2.3$~ks, respectively. For the reduction and
analysis of the data from the three instruments onboard the
\emph{Swift} satellite, we used the \texttt{HEASoft v. 6.4} package, 
together with the \texttt{CALDB} updated on February $13$, $2008$.

The data from the BAT instrument \citep{BAT}, optimized for the $15-195$~keV 
energy band, were binned, cleaned of hot pixels, and deconvolved. 
The intensity maps from the individual pointings were integrated by using 
the corresponding variance maps as a weighting factor, and the resulting 
exposure simultaneous to the XRT and UVOT observations was $5.4$~ks. 
No source was detected; the corresponding $3\sigma$ upper limits were 
$2.7\times 10^{-10}$~erg~cm$^{-2}$~s$^{-1}$ and 
$4.3\times 10^{-10}$~erg~cm$^{-2}$~s$^{-1}$ in the $20-40$~keV and 
$40-100$~keV energy band, respectively. 
However, the blazar was detected in the $9$-month data survey \citep{Tueller}, 
and the source spectrum is publicly available online \citep{BATSURVEY}. 
The fitting of a power-law model is optimal for a photon index of 
$\Gamma=1.5\pm 0.4$ and a flux of $4.8\times 10^{-11}$~erg~cm$^{-2}$~s$^{-1}$ 
in the $20-100$~keV energy band. 

The X-Ray Telescope (XRT, Burrows et al.~2005), operating in the $0.2-10$~keV energy 
band, was set to work in photon counting mode during the two pointings 
on PKS~$2149-306$. 
Data were processed and screened by using the \texttt{xrtpipeline} task 
with the single to quadruple pixels events selected (grades $0-12$). 

The spectral analysis of the XRT data was presented in 
Sambruna et al.~(2007) as a joint fit of the two observations. 
Here, we consider the two data sets separately and we fit each with three 
models, all including Galactic extinction: a power-law (\texttt{wabs*zpo}), a 
power-law with an extra-absorption component in the source rest frame 
(\texttt{wabs*zwabs*zpo}) and a broken power-law (\texttt{wabs*bknpo}). 

The spectrum of the first observation ($2005$-$12$-$10$) is well described by a
single power-law with Galactic absorption (Table~\ref{fits}). 
Residuals are shown in panel \textit{b} of Fig.~\ref{res} (in units of 
$1~ \sigma$). 
Although the broken power-law and the power-law with extra absorption models 
yield $\chi^2_r=1.2$ (for $27$ and $28$ dof, respectively), the spectral 
parameters are not well constrained or they converge to unphysical values. 
For the second observation ($2005$-$12$-$13$), we found that the broken
power-law function provided the best-fit model, although both the power-law 
model and the power-law function with an extra-absorption component described 
the observational data accurately. 
In Table~{\ref{fits} we present the best-fit parameters for the three applied 
models. 
The residual plot of the power-law model (Fig.~\ref{res}, panel \textit{c})
shows a hint of curvature, which is not present when the 
broken power-law model is assumed (Fig.~\ref{res}, panel \textit{d}). 
The curved model is also supported by the F-test analysis, showing that the
flatter power-law function below $\sim 1.7$~keV is required with a 
probability of $98.6\%$. 
The quality of data acquired to date does not allow us to distinguish firmly 
between the intrinsically curvature (broken power-law) and the 
extra-absorption model. 

UVOT \citep{UVOT} data analysis was performed by the 
\texttt{uvotmaghist} task using  a source region of $5$'' radius for the 
optical and $10$'' for the UV filters.
Since a contaminating source is located close to PKS~$2149-306$, the 
background was evaluated from a nearby region of $60$'' radius. 
The observed magnitudes were computed as the mean values of recorded data, 
since they do not exhibit significant variations: 
$V_{543\rm{nm}}=17.4 \pm 0.2$,
$B_{434\rm{nm}}=17.7 \pm 0.1$, $U_{344\rm{nm}}=17.1 \pm 0.1$,
$UVW1_{291\rm{nm}} = 17.9 \pm 0.2$, $UVM2_{231\rm{nm}} \ge 19.1$, and 
$UVW2_{212\rm{nm}} \ge 19.4$. 
The optical magnitudes are in accordance with values observed by 
Francis et al.~(2000). 
With respect to the previous analysis of Sambruna et al.~(2007), we obtain 
comparable magnitudes for all filters but $UVM2$ and $UVW2$, for which we 
found an upper limit. 
The discrepancy is probably due to the use of updated calibration files. 

\subsection{Joint fits} \label{joint}
We attempted a joint fit of the spectra provided by EPIC and ISGRI data. 
This is justified by the fact that the source shows a low flux level in both 
ISGRI and EPIC data and that the ISGRI data fall on the extrapolation of the 
EPIC spectrum (the normalization of the ISGRI spectrum with respect to the 
EPIC spectrum is $C=1.0 \pm 0.4$). 
We then argue that the blazar is in the same spectral state during the two 
observations, despite being separated by $\sim 4$ years. 
The joint fitting procedure found a single power-law function to be the 
best-fit model with parameters given in Table~\ref{fits}. 

We performed a joint fit to the XRT observations and the BAT spectrum. 
Best-fit models are given in Table~\ref{fits}. 
We note that for the observation of $2005$-$12$-$13$ the normalization of BAT 
with respect to XRT is higher than unity ($C=2.3^{+2.2}_{-1.0}$): this 
probably reflects the flux variability of the source, given the fact that the 
BAT spectrum relates to data integrated over a period of 9 months. 
Since the XRT spectrum was retained as reference, we note that the model flux 
is comparable to that obtained for the EPIC+ISGRI spectrum. 

\section{Discussion}
\begin{table*}[!t]
\begin{minipage}[t]{\textwidth}
\caption{\label{fits} Fit results.}
\begin{center}
\renewcommand{\footnoterule}{}
\begin{tabular}{lcccccc}
\hline
\hline
\multicolumn{7}{c}{X-rays ($0.2-10$~keV)}\\
\hline
\hline
Instrument &  Model & $\Gamma$ or $\Gamma_1$ & \multicolumn{2}{c}{Additional Parameters} & $F_{\rm 2-10~keV}$ & $\chi^2_r$/dof \\
{}         &        & {}                     &                      &                    &                    &                \\
\hline
\emph{XMM}/EPIC           & \texttt{wabs*zpo}       & $1.42 \pm 0.01$ & {}  & {} & $0.56 \pm 0.01$ & $1.08/977$\\
\emph{Swift}/XRT (OBS.~1) & \texttt{wabs*zpo}       & $1.44\pm 0.07$  & {}  & {} & $1.2 \pm 0.1$   & $1.24/29$ \\
\emph{Swift}/XRT (OBS.~2) & \texttt{wabs*zpo}       & $1.38 \pm 0.09$ &     &    & $1.04^{+0.06}_{-0.09}$  & $1.63/15$ \\
                          & \texttt{wabs*zwabs*zpo} & $1.55^{+0.17}_{-0.16}$& $N_z=1.0^{+1.1}_{-0.8}$~$10^{22}$~cm$^{-2}$ &     & $0.97^{+0.5}_{-0.08}$& $1.42$/$14$ \\
                          & \texttt{wabs*bknpo}     & $0.98_{-0.41}^{+0.29}$& $E_{\rm break}= 1.9_{-0.6}^{+1.5}$~keV & $\Gamma_2=1.86_{-0.31}^{+1.19}$ & $0.8 \pm 0.1$ & $0.98/13$ \\
{} \\
\hline
\hline
\multicolumn{7}{c}{Hard X-rays ($20-100$~keV)}\\
\hline
\hline
Instrument            & Model & $\Gamma$ & {} & {} & $F_{\rm20-100~keV}$                & $\chi^2_r$/dof \\
{}                    & {} & {}       & {} & {} &     & {}\\
\hline
\emph{INTEGRAL}/ISGRI & \texttt{zpo} & $1.5_{-0.8}^{+0.9}$ & {} & {} & $2.1 \pm 0.2$ & $0.51/3$\\
\emph{Swift}/BAT      & \texttt{zpo} & $1.5\pm 0.4$        & {} & {} & $4.8 \pm 0.5$ & $0.48/2$\\
{} \\
\hline
\hline
\multicolumn{7}{c}{Joint fit ($0.2 -100$~keV)}\\
\hline
\hline
Instrument & Model & $\Gamma$ or $\Gamma_1$ & \multicolumn{2}{c}{Additional Parameters}  & $F_{\rm2-100~keV}$ & $\chi^2_r$/dof \\
\hline
EPIC+ISGRI   & \texttt{wabs*zpo}   & $1.42 \pm 0.01$ &             &                     & $3.13^{+0.05}_{-0.04}$       & $1.08/980$ \\
XRT (OBS.~1) + BAT & \texttt{wabs*zpo}   & $1.44 \pm 0.07$ & & & $6.4^{+0.9}_{-0.8}$ & $1.2/32 $\\
XRT (OBS.~2) + BAT & \texttt{wabs*bknpo} & $0.9^{+0.3}_{-0.4} $   & $E_{\rm break}=1.6^{+0.9}_{-0.4}$~keV & $\Gamma_2=1.7^{+0.3}_{-0.2}$ & $3 \pm 1$ & $0.92/16$ \\

\end{tabular}
\end{center}
\footnotesize{Models and parameters for single instruments and joint fit. Applied models are: a power-law in the source rest frame (\texttt{zpo}), a power-law with an absorption component in the object rest frame (\texttt{zwabs*zpo}) and a broken power-law (\texttt{bknpo}). All models include Galactic absorption (\texttt{wabs}) with column density $N_{\rm H}=1.63\times 10^{20}$~cm$^{-2}$ \citep{LAB}. Flux is given in units of $10^{-11}$~erg~cm$^{-2}$~s$^{-1}$.}
\end{minipage}
\end{table*}

\begin{figure}
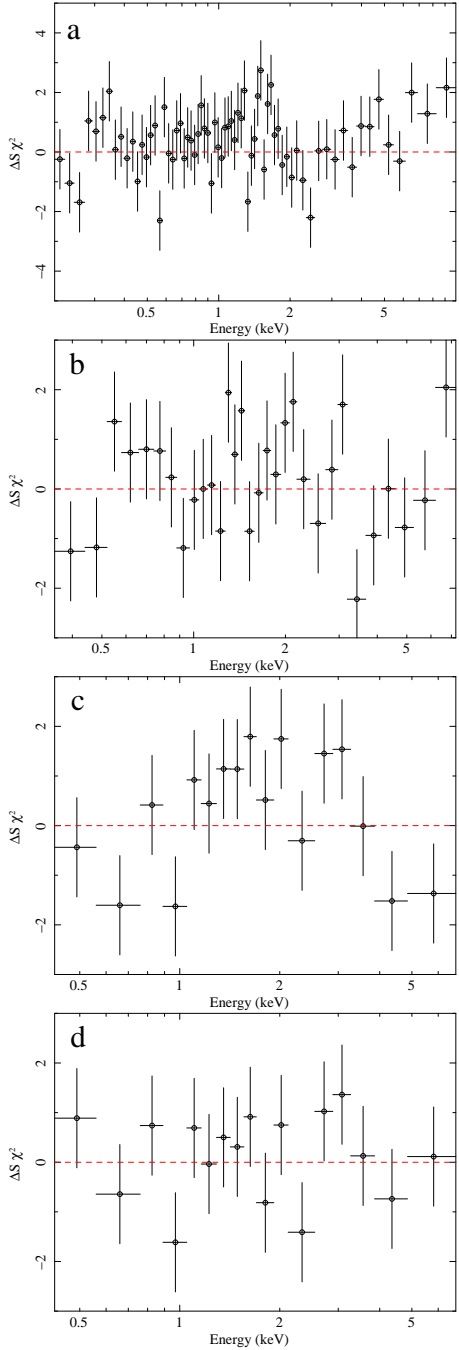

\centering
\includegraphics[angle=-90,scale=0.25]{1128f1a.ps}
\includegraphics[angle=-90,scale=0.25]{1128f1b.ps}
\includegraphics[angle=-90,scale=0.25]{1128f1c.ps}
\includegraphics[angle=-90,scale=0.25]{1128f1d.ps}
\caption{\label{res} Residual plots in units of $\sigma$ with error bars of 
size $1$ sigma. \textit{Panel a}: \emph{XMM-Newton} data with respect to a 
power-law model in the object rest frame with Galactic absorption; 
\textit{Panel b}: XRT~(2005-12-10) with a single power-law; \textit{Panel c}: 
XRT~(2005-12-13) with a single power-law; \textit{Panel d}: XRT~(2005-12-13) 
using a broken power-law model.}
\end{figure}

\subsection{Spectral variation}
The present analysis, and past results, shows that the hard-X-ray emission in 
the $\sim 60-300$~keV interval (object rest frame) of PKS~$2149-306$ does 
not vary in spectral shape: indeed spectra obtained almost a year apart by 
\emph{INTEGRAL}/ISGRI and \emph{Swift}/BAT show a constant (within errors) 
photon index, which is also consistent with the slope previously inferred for 
data acquired by \emph{BeppoSAX} in $1997$ ($\Gamma = 1.37 \pm 0.04$ - 
Elvis et al.~2000). 
We recall that the spectra refer to an average state covering $\sim 10$~days 
for ISGRI and $\sim 9$~months for BAT. 
Despite the lack of spectral variability, the integrated flux in the 
$20-100$~keV band exhibits a variation of more than a factor of two from the 
lowest value of ISGRI to the highest value for BAT. 
A remarkable change in the hard-X-ray flux is not uncommon for blazars, and it 
was observed, for instance, in \emph{Swift}~J$1656.3-3302$, although with a 
different spectral slope \citep{MASETTI}. 

In the $0.2-10$~keV energy band, our data analysis suggests a more complex
behavior. 
As discussed in Ferrero \& Brinkmann~(2003), the source spectrum acquired in $2001$
(\emph{XMM-Newton}/EPIC data) is well described by a single power-law function 
with little evidence that a more complex model is necessary. 
Instead, the two observations with XRT in $2005$ suggest a spectral change 
from a power-law model in the first observation ($2005$-$12$-$10$), to a 
curved model in the second pointing ($2005$-$12$-$13$), with a hint of 
curvature developing over $\sim 3$~days. 
As for other high-z radio-loud quasars, this feature is usually interpreted as 
an extra-absorption component. 
This absorbing cloud should be associated with the quasar frame, since no 
Damped Ly$\alpha$ features reveal absorption by intervening material. 
Although short-term variability due to changes in the intrinsic absorption was observed in Seyfert galaxies 
\citep{RISALITI07}, a variation on time intervals of days is difficult to reconcile 
with the blazar picture since the absorbing cloud should be located at a 
distance of $\sim 10^{15}$~cm from the black hole, below the length scale on 
which jet dissipation is supposed to occur in this class of objects. 
On the other hand, in the blazar scenario the soft photon deficit and its variation can be naturally 
interpreted as a spectral break due to the Inverse Compton regime onset 
\citep{TAVECCHIO07,SAMBRUNA07}. 
Indeed a low-energy roll-off appears to be a common property of FSRQs, for 
distant objects such as MG$3$~J$225155+2217$ ($z=3.668$, 
Maraschi et al.~2008, RBS~$315$ ($z=2.69$, Tavecchio et al.~2007), and 
nearby FSRQs such as $3$C~$454.3$ ($z=0.859$, Ghisellini et al.~2007). 
For high-z FSRQs, the break signature falls in the soft X-ray band, where this 
feature can be blurred by contribution from the primary seed photon input and 
the bulk Comptonization of cold electrons in the jet \citep{CELOTTI07}. 
Apart from spectral analysis, a variability study of simultaneous data in the 
optical to X-ray bands can provide clues to the origin of the intrinsic 
curvature in the continuum \citep{FOSCHINI08}.

\subsection{SED modeling}
\begin{figure}
\includegraphics[scale=0.45]{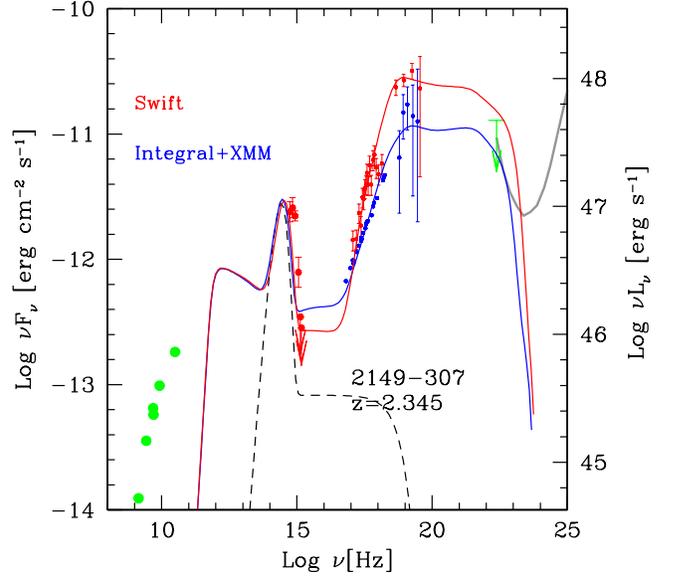}
\caption{\label{sed} Spectral Energy Distributions and models (solid
lines) for \emph{Swift} (red) and \emph{XMM-Newton}+ISGRI (blue) data. The
figure shows in green archival radio and IR data (from NED) and the EGRET 
upper limit \citep{FITCHEL94}. 
The gray line represents the \emph{GLAST} sensitivity curve. 
The dashed line shows the input spectrum from the disk and the corona.}
\end{figure}

\begin{table}[!t]
\begin{center}
\caption{\label{parsed} SED Model Parameters. See text for more details.}
\begin{tabular}{lrr}
\hline
\hline
Parameter & \emph{Swift}  & EPIC+ISGRI \\
$R(10^{15}\rm{cm})$       & 15   &  15  \\
$h(10^{15}\rm{cm})$       & 150  & 150  \\
$\Theta$(deg)             & 3    &   3  \\
$\Gamma_{bulk}$           & 15   &  10  \\
$L_{inj}(10^{43}\rm{erg} \, \rm{s}^{-1})$ & 8 & 8 \\
$\gamma_{break}$          & 5    &   5  \\
$\gamma_{max}$            & 1000 & 1000 \\
$B$(G)                    & 3.3  &   3.3\\
$L_{disk}(10^{45}\rm{erg} \, \rm{s}^{-1})$ & 150  & 150 \\
$R_{BLR}(10^{15}\rm{cm})$ & 1300 & 1300 \\
\hline
\hline
\end{tabular}
\end{center}
\end{table}

We define two different states of PKS~$2149-306$: a ``low-flux'' state 
given by \emph{XMM-Newton} and \emph{INTEGRAL} points and a ``high-flux'' 
state studied with XRT and BAT data, for which we obtained a high flux level 
separately in the $[2-10]$ and $[20-100]$~keV band. 
We chose the second observation of XRT (on $2005$-$12$-$13$), since the flux 
is comparable to the first one but exhibits spectral bending at low energies. 
In Fig.\ref{sed} the SEDs for both states are depicted separately in blue 
and red. 
Optical/UV data were corrected for Galactic absorption according to 
Cardelli et al.~(1989) and using the zero points given in the calibration files. 
X-ray points are corrected only for the Galactic absorption.  
The multiwavelength spectra are completed with radio and IR data taken from 
NED, and the EGRET upper limit \citep{FITCHEL94}. 

We attempt to describe the two states in terms of changes in the physical 
parameters of the source. 
For both, we model the source as discussed in detail in Ghisellini et al.~(2002). 
The model assumes synchrotron and Inverse Compton emission from a spherical 
region in the jet, of radius $R$, at a distance $h$ from the disk, with 
bulk Lorentz factor $\Gamma$ and orientation $\Theta$ with respect to the line 
of sight. 
The blob is populated by a tangled and homogeneous magnetic field $B$ and by 
relativistic electrons, injected with total power $L_{inj}$ and a broken 
power-law energy distribution described by $\gamma^{-1}$ in 
$[1,\gamma_{break}]$, and $\gamma^{-3.2}$ up to $\gamma_{max}$. 
The model accounts for radiative losses and determines the cooling energy 
$\gamma_{cool}$ of particles after a dynamical time $R/c$ 
(see e.g., Maraschi et al.~2008). 
The primary emission is provided by a disk and a tenuous corona with a 
power-law spectrum of slope $\alpha =1$ and cut-off at $150$~keV.
The disk spectrum is described by a black-body peaking at 
$\nu_0 \sim 10^{15}$~Hz and with power $L_{disk}$, derived from the 
optical/UV spectral data. 
For the \emph{XMM-Newton}+\emph{INTEGRAL} state, we can place only weak constraints 
on the disk/corona emission, because of the lack of data from the optical 
monitors of both satellites. 
However, the upper limit of \emph{XMM-Newton}/OM in the UV band and the almost 
constant archival optical data suggest that the soft photon input for the 
Inverse Compton regime does not vary significantly in the two states. 
The Broad Line Region (BLR), located at $R_{BLR}$ from the central 
black hole (see e.g., Ghisellini et al.~1996), reflects the primary 
emission toward the jet with luminosity $L_{BLR}=0.1 L_{disk}$ 
\citep{KASPI05}. 

We attempted to reproduce both states by varying as small a number of 
parameters as possible. 
This choice was also motivated by the limited energy range covered by the 
available simultaneous data. 
The absence of information above hundreds of keV makes the high energy 
electron distribution poorly determined. 
The synchrotron regime is also poorly constrained because of the 
lack of data. This is, however, unimportant for the SED modeling, since the radio-IR emission
includes contributions from the outer regions of the jet and, therefore, is not considered in this model.
 
Input model parameters for both states are given in Table~\ref{parsed}. 
The overall SED of PKS~$2149-306$ is typical of FSRQs at high redshift, with 
the Inverse Compton mechanism dominating the spectrum from X- to 
$\gamma$-rays and in accordance with the standard blazar sequence 
\citep{MARASCHI08}.

The two different SEDs can be explained in terms of a change only in the jet 
bulk Lorentz factor, being $\Gamma=15$ for the ``high-flux'' state and 
$\Gamma=10$ for the ``low-flux'' state. 
The different Lorentz factors of the emitting region can be interpreted in 
terms of the inner-shock model scenario (e.g., Spada et al.~2001), which 
assumes that subsequent colliding shells at different speed produce a new 
radiating shell with intermediate $\Gamma$. 
The variation in the $\Gamma$ Lorentz factor accounts for the change in the 
total power of the source and for the spectral changes in the Inverse Compton 
regime, observed in the SEDs for the two states. 
In the ``low-flux state'' (found in \emph{XMM-Newton} and \emph{INTEGRAL} 
data), the ``low'' $\Gamma$ drives a ``cooling branch'' in the electron 
distribution with $\gamma_{cool}=12$ and a slope $2.2$ in 
$[\gamma_{peak},\gamma_{cool}]$. 
This cooling branch does not develop in the ``high-flux'' state, 
since we found that $\gamma_{cool} \sim \gamma_{break}$. 

The present data quality and the partial simultaneous spectral coverage 
limit a careful inspection of all full model input parameter range. 
However, it is remarkable that the two spectral states are described by 
changing only the jet bulk Lorentz factor; this parameter accounts for both 
the flux variation and the spectral change of the Inverse Compton spectra. 
More complex solutions, changing more than one parameter, are not excluded, 
although we propose the simplest and physically consistent explanation of the 
state variation for the source.

\section{Conclusions}
We have investigated the long-term behavior of the blazar PKS~$2149-306$ 
($z=2.345$). 
All \emph{XMM-Newton}, \emph{Swift}, and \emph{INTEGRAL} publicly available 
data have been reprocessed and analyzed with the latest software versions and 
calibration data bases. 
The SED, compiled with the data analyzed in the present work complemented with 
those available in the archives, is typical of FSRQs at the brightest end of 
the blazar sequence. 

\emph{Swift} observations suggest a hint of a roll-off below $\sim 1$~keV 
emerging on $\sim 3$-day timescale, which can be explained as the natural 
spectral break due to the Inverse Compton onset. 
However, the present data quality does not allow us to distinguish firmly 
between the absorption/curvature dichotomy, and further multiband and 
simultaneous observations are necessary. 

In the $60-300$~keV energy band (in the blazar frame), the flux levels derived 
from ISGRI and BAT data differ by more than a factor of two, without spectral 
slope variation. 
From the broad-band spectral analysis, we can identify two different states: 
one found in data until $2004$ (\emph{XMM-Newton} and \emph{INTEGRAL}), and 
another referring to a \emph{Swift} observation performed in $2005$. 
The SED modeling shows that the two states can be reproduced by changing 
only the bulk Lorentz factor of the jet.

\begin{acknowledgements}
This research has made use of data obtained from the High Energy
Astrophysics Science Archive Research Center (HEASARC), provided by
NASA's Goddard Space Flight Center. 
This research has made use of the NASA/IPAC Extragalactic Database (NED) 
which is operated by the Jet Propulsion Laboratory, California Institute of 
Technology, under contract with the National Aeronautics and Space 
Administration. 
We acknowledge the use of public data from the Swift data archive. 
We acknowledge partial support from ASI/INAF Contract I/088/06/0.
\end{acknowledgements}

\bibliographystyle{aa}

\begin{thebibliography}{}

\bibitem[Barthelmy et al.~2005]{BAT} Barthelmy,~S.~D., Barbier,~L.M., 
  Cummings,~J.M., et al. 2005, Space Sci. Rev. 120, 143

\bibitem[Baumgartner et al.~2008]{BATSURVEY} Baumgartner,~W., Tueller,~J., 
  Mushotzky,~R., et al. 2008, ATel 1429

\bibitem[Bird et al.~2008]{BIRD} Bird,~A.J., Malizia,~A., Bazzano,~A., et al. 
  2007, ApJS 170, 175

\bibitem[Burrows et al.~2005]{XRT} Burrows,~D.N., Hill,~J.E., Nousek,~J.A., 
  et al. 2005, Space Sci. Rev. 120, 165

\bibitem[Cappi et al.~1997]{CAPPI97} Cappi,~M., Matsuoka,~M., Comastri,~A., 
  et al. 1997, ApJ 478, 492

\bibitem[Cardelli et al.~1989]{CARDELLI89} Cardelli,~J.~A., Clayton,~G.~C. \& 
  Mathis,~J.~S. 1989, ApJ 345, 245

\bibitem[Celotti et al.~2007]{CELOTTI07} Celotti,~A., Ghisellini,~G. \&
  Fabian,~A.C. 2007, MNRAS 375, 417

\bibitem[Courvoisier et al.~2003]{COURVOISIER} Courvoisier,~T.~J.~L., 
  Walter,~R., Beckmann,~V., et al. 2003, A\&A 411, L53

\bibitem[Elvis et al.~2000]{ELVIS00} Elvis,~M., Fiore,~F., Siemiginowska,~A., 
  et al. 2000, ApJ 543, 545

\bibitem[Fang et al.~2001]{FANG01} Fang,~T., Marshall,~H.~L., Bryan,~G.~L. \& 
  Canizares,~C.~R. 2001, ApJ 555, 356

\bibitem[Ferrero \& Brinkmann~2003]{FERRERO03} Ferrero,~E. \& Brinkmann,~W. 
  2003, A\&A 402, 465

\bibitem[Fichtel et al.~1994]{FITCHEL94} Fichtel,~C.~E., Bertsch,~D.~L., 
  Chiang,~J., et al. 1994, ApJSS 94, 551

\bibitem[Foschini et al.~2006]{FOSCHINI06} Foschini,~L., Ghisellini,~G., 
  Raiteri,~C.~M., et al. 2006, A\&A 453, 829

\bibitem[Foschini~2008]{FOSCHINI08} Foschini,~L. 2008, Adv. Space Res., 
  accepted for publication, [\texttt{arXiv:0807.2253v1}]

\bibitem[Francis et al.~2000]{FRANCIS00} Francis,~P.~J., Whiting,~M.~T. \& Webster,~R.~L. 2000, PASA 53, 56 

\bibitem[Ghisellini \& Madau~1996]{GHISELLINI96} Ghisellini,~G.\& Madau,~P. 
  1996, MNRAS, 280, 67

\bibitem[Ghisellini et al.~2002]{GHISELLINI02} 
  Ghisellini,~G., Celotti,~A. \& Costamante,~L. 2002, A\&A 386, 842

\bibitem[Ghisellini et al.~2007]{GHISELLINI07} Ghisellini,~G., Foschini,~L., 
  Tavecchio,~F. \& Pian,~ E. 2007, MNRAS 382, L82

\bibitem[Goldwurm et al.~2003]{GOLDWURM03} Goldwurm,~A., David,~P., 
  Foschini,~L., et al. 2003, A\&A 411, L223

\bibitem[Kalberla et al.~2005]{LAB} Kalberla,~P.~M.~W., Burton,~W.~B., 
  Hartmann,~D., et al. 2005, A\&A 440, 775

\bibitem[Kaspi et al.~2005]{KASPI05} Kaspi,~S., Maoz,~D., Netzer,~H., 
  et al. 2005, ApJ, 629, 61

\bibitem[Lebrun et al.~2003]{LEBRUN03} Lebrun,~F., Leray,~J.~P.,Lavocat,~P. 
  et al. 2003, A\&A 411, L141

\bibitem[Maraschi et al.~2008]{MARASCHI08} Maraschi,~L., Foschini,~L., 
  Ghisellini,~G., et al. 2008, MNRAS, 391, 1981

\bibitem[Masetti et al.~2007]{MASETTI} Masetti,~N., Mason,~E., Landi,~R., 
  et al. 2008, A\&A 480, 715

\bibitem[Mason et al.~2001]{OM} Mason,~K.~O., Breeveld,~A., Much,~R., et al. 
  2001, A\&A 365, L36

\bibitem[Ojha et al.~2005]{OJHA05} Ojha,~R., Fey,~A.~L., Charlot,~P., et al. 
  2005, AJ 130, 2529

\bibitem[Page et al.~2004]{PAGE04} Page,~K.~L., O'Brien,~P.~T., Reeves,~J.~N. 
  \& Turner,~M.~J.~L. 2004, MNRAS 347, 316

\bibitem[Page et al.~2005]{PAGE05} Page,~K.~L., Reeves,~J.~N., O'Brien,~P.~T. 
  \& Turner,~M.~J.~L. 2005, MNRAS 364, 195

\bibitem[Risaliti et al.~2007]{RISALITI07} Risaliti,~G., Elvis,~M., \& 
  Fabbiano,~G., et al. 2007, ApJ 659, L111

\bibitem[Roming et al.~2005]{UVOT} Roming,~P.~W.~A., Kennedy,~T.~E., 
  Mason,~K.~O., et al. 2005, Space Sci. Rev. 120, 95

\bibitem[Sambruna et al.~2007]{SAMBRUNA07} Sambruna,~R.~M., Tavecchio,~F., 
  Ghisellini,~G., et al. 2007, ApJ 669, 884

\bibitem[Siebert et al.~1996]{SIEBERT96} Siebert,~J., Matsuoka,~M., 
  Brinkmann,~W., et al. 1996, A\&A 307, 8

\bibitem[Spada et al.~2001]{SPADA01} Spada,~M., Ghisellini,~G., Lazzati,~D. \& 
  Celotti,~A. 2001, MNRAS 325, 1559

\bibitem[Str\"uder et al.~2001]{EPICPN} Str\"uder,~L., Briel,~U., 
  Dennerl,~K., et al. 2001, A\&A, 365, L18

\bibitem[Tavecchio et al.~2007]{TAVECCHIO07} Tavecchio,~F., Maraschi,~L., 
  Ghisellini,~G., et al. 2007, ApJ 665, 980

\bibitem[Tueller et al.~2008]{Tueller} Tueller,~J., Mushotzky,~R.~F., 
  Barthelmy,~S., et al. 2008, ApJ 681, 113

\bibitem[Turner et al.~2001]{MOS} Turner,~M.~J., Abbey,~A., Arnaud,~M., 
  et al. 2001, A\&A, 365, L27

\bibitem[Ubertini et al.~2003]{UBERTINI03} Ubertini~P., Lebrun~F., 
  Di~Cocco~G., et al. 2003, A\&A 411, L131

\bibitem[Yaqoob et al.~1999]{YAQOOB99} Yaqoob,~T., George,~I.~M., Nandra,~K. 
  et al. 1999, ApJ 525, L9

\bibitem[Wilkes~1986]{bw86} Wilkes,~B.~J. 1986, MNRAS 218, 331

\end{thebibliography}

\end{document}